\documentclass[prb,
twocolumn,
 amsmath,amssymb,
 aps,
]{revtex4-2}

\usepackage{graphicx}
\usepackage{dcolumn}
\usepackage{bm}
\usepackage{braket}

\usepackage{hyperref}
\hypersetup{colorlinks,urlcolor=blue,linkcolor=red,          
citecolor=blue,        
filecolor=blue}

\begin{document}

\title{Fate of entanglement in open quantum spin liquid: Time evolution of its genuine multipartite negativity upon sudden coupling to a dissipative bosonic environment}

\author{Federico Garcia-Gaitan}
\author{Branislav K. Nikoli\'c}
\email{bnikolic@udel.edu}
\affiliation{Department of Physics and Astronomy, University of Delaware, Newark, DE 19716, USA}

\begin{abstract}
    Many-body entanglement properties of quantum spin liquids (QSLs), persisting at arbitrarily long distances, have been intensely explored over the past two decades, but mostly for QSLs viewed as {\em closed}  quantum systems. However, in experiments and potential quantum computing applications, candidate materials for this exotic phase of quantum matter will always interact with a dissipative environment, such as the one generated by bosonic quasiparticles in solids at finite temperature. Here we investigate both the {\em stability} and {\em spatial distribution}  of entanglement for the Kitaev model of QSL, which is made {\em open} by its sudden coupling to an infinite bosonic bath of Caldeira-Leggett type and then time-evolved in both Markovian and non-Markovian regimes. From the time-dependent density matrix of QSL subregions, we extract genuine multipartite negativity (GMN), quantum Fisher information, spin-spin correlators, and the expectation value (EV) of the Wilson loop operator. In particular, time dependence of GMN offers the most penetrating insights: ({\em i}) in the Markovian regime, it remains nonzero only in hexagonal loopy subregions of QSL (as also discovered very recently for closed QSLs), eventually vanishing on the same timescale on which the EV of the  Wilson loop operator vanishes; ({\em ii})  in the non-Markovian regime with pronounced memory effects, surprisingly, GMN remains nonzero up to much higher temperatures while also remaining zero in non-loopy subregions. In addition, the non-Markovian dynamics generates emergent  interactions between spins, thereby opening avenues for tailoring properties of QSL via engineering of dissipation.
\end{abstract}

\maketitle

\section{Introduction}\label{sec:intro}

Quantum spin liquids (QSLs)~\cite{Broholm2020,Savary2017, Knolle2019} are exotic phases of matter that, despite being composed of magnetic atoms, do not exhibit long-range magnetic ordering down to absolute zero temperature. Instead, they can exhibit  topological ground-state degeneracy, long-range entanglement~\cite{Jiang2012a,Grover2013,Semeghini2021,Savary2017,Knolle2019}, excitations with fractional quantum statistics~\cite{Knolle2015,Do2017,Nasu2023,Harada2024,Zhu2019c,Joy2025,Chen2025b} and emergent gauge fields~\cite{Sachdev2023}. Due to fundamental interest, as well as potential applications in quantum computing robust against decoherence~\cite{Kitaev2003,Nayak2008,Semeghini2021}, materials hosting QSLs are highly sought~\cite{Broholm2020}. However, despite  many candidates and diverse experimental probes~\cite{Banerjee2016, Banerjee2017,Balz2019,Kasahara2022} applied to them, definitive proof for the existence of the QSL phase remains elusive~\cite{Winter2017,Maksimov2020,Kasahara2022, Matsuda2025} (note that the  QSL phase has been demonstrated on  programmable quantum simulators~\cite{Semeghini2021, Evered2025a, Park2025}). 

Since QSLs cannot be characterized by traditional local order parameters and broken symmetry~\cite{Sachdev2023}, entanglement entropies (EEs) and their linear combinations have been introduced~\cite{Kitaev2006a,Levin2006} and intensely (typically numerically) studied for gapped~\cite{Jiang2012a,Depenbrock2012,Grover2013,Wildeboer2017} and gapless~\cite{Swingle2012,Grover2013,Zhu2018a} QSLs. They quantify {\em bipartite}  entanglement between a subregion of QSL and its complement. In particular, one such combination yields topological entanglement entropy (TEE)~\cite{Kitaev2006a, Levin2006,Kim2023,Levin2024} that has been considered  a ``smoking gun''~\cite{Jiang2012a} evidence of topological order 
in the {\em pure} ground state $|\mathrm{GS}\rangle$ of gapped QSL considered as a {\em closed} quantum system.  However, TEE can fail to differentiate topological phases with distinct fractionalized excitations (for which TEE might be identical~\cite{Flammia2009,Bridgeman2016,Wildeboer2017,Liu2024c}); or it can produce spurious signals (such as nonzero TEE for topologically trivial GS~\cite{Zou2016}). Thus, a finer probe could be offered by {\em multipartite} entanglement measures~\cite{Liu2024c}. The archetypical example of a pure state with maximal multipartite entanglement is the Greenberger–Horne–Zeilinger state of $N$ spins $S=1/2$ or qubits, $|\mathrm{GHZ}\rangle = (|\!\!\uparrow\rangle^{\otimes N} + |\!\!\downarrow\rangle^{\otimes N})/\sqrt{2}$,  with zero entanglement between any pair of $N-1$ spins once one spin is traced out.  A particularly interesting result on the multipartite nature of entanglement, as well as its {\em spatial distribution}, in closed QSLs has been obtained~\cite{Lyu2025} very recently by analyzing the so-called genuine multipartite negativity (GMN)~\cite{Jungnitsch2011,Hofmann2014, Song2025,Wang2025a,Sabharwal2025} between $m \le 6$ parties, where each party is a single spin or a group of spins. GMN employs the partial transpose operation used to detect entanglement of mixed quantum states via more familiar entanglement   negativity~\cite{Peres1996, Lu2020, Lu2023, Fan2024}, to which GMN reduces for $m=2$ parties. It has revealed that multipartite entanglement between spins is {\em absent} in the small non-loopy subregions of QSLs, while being nonzero for loop-shaped (such as hexagons in Kitaev QSL or ``bow–tie'' in kagome antiferromagnet~\cite{Lyu2025,Sabharwal2025}) larger subregions. Since GMN quantifies the collective entanglement between remote degrees of freedom, its application to QSLs could deepen our  understanding of fractionalization and  encoding of quantum information by gauge bosons.

There has also been a growing interest in finding proper quantities for diagnosing topological order and/or long-range entanglement in  {\em mixed} quantum states, such as Gibbs states~\cite{Trushechkin2022,Nathan2024} generated by thermal fluctuations~\cite{Lu2020, Lu2023,Sabharwal2025,Zhou2025} or those generated by decoherence (occurring even when energy-exchange processes are suppressed~\cite{Joos2003,Merkli2007}) and dephasing~\cite{Fan2024, Lee2023, Ellison2025, Sohal2025, Wang2025} as major obstacles in present noisy intermediate-scale quantum devices. However, these efforts do not consider possibly structured~\cite{Fux2023,Fux2024,Cygorek2024,Cygorek2024a}   environments and thereby induced dissipative dynamics to which candidate materials are always exposed, typically due to bosonic quasiparticles in solids. Beyond the  ultraweak system-bath coupling limit, such  environments can lead to time-asymptotic steady states that are thermalized~\cite{Fux2023} but, however, {\em not} of Gibbs type~\cite{Fux2023,Trushechkin2022,Lee2022,TelloBreuer2024};  and they can also  induce~\cite{ReyesOsorio2026} new effective interactions between spins.

Although {\em open} quantum system approaches tailored for QSLs have emerged~\cite{Yang2021,Yang2022,Kulkarni2022,Shackleton2024,Harada2024,Hwang2024,Coser2019,Fukui2024}, they typically assume simplified phenomenological models of the environment (i.e., without starting from the environment Hamiltonian and then tracing its microscopic degrees of freedom). They also focus almost {\em exclusively} on the Markovian regime (i.e., when system-environment coupling is weak and environment correlations are short compared to the timescale of the system evolution~\cite{Rudner2020, Fux2023}) describing it by the Lindblad quantum master equation (QME)~\cite{Lindblad1976,Manzano2020,Rudner2020, Nathan2024}. Since numerical time evolution of the density matrix via Lindblad QME becomes computationally expensive for many spins~\cite{Pocklington2025}, most of such studies try to evade solving it altogether and instead focus on analyzing spectral properties of an effective non-Hermitian Hamiltonian (composed of the original QSL Hamiltonian plus the Lindblad operators). This precludes understanding of the fate of entanglement, as its quantification requires a time-evolved density matrix. Furthermore, the {\em non-Markovian} dynamics with pronounced memory effect~\cite{Breuer2016,Vega2017} of open QSLs, which is particularly relevant~\cite{Gulacsi2023} in quantum computing applications, remains unaddressed. This lack of studies of non-Markovian open QSLs is closely related to the fact that formalisms for the time evolution of {\em many} spins strongly coupled to a dissipative bosonic environment have become available only very recently~\cite{Fux2023, Fux2024, Cygorek2024a, Cygorek2024, ReyesOsorio2026}.

In this study, we investigate how robust the entanglement of QSL is via real-time evolution of its reduced density matrix $\hat{\rho}(t)$ in {\em both} Markovian and non-Markovian regimes. For this purpose, we utilize the universal Lindblad QME~\cite{Rudner2020, Nathan2024} in the former regime and tensor network (TN) methodologies~\cite{Fux2023, Fux2024,Cygorek2024a, Cygorek2024} for open quantum systems or the reaction coordinate (RC)~\cite{Schaller2014} + polaron~\cite{Anto-Sztrikacs2023,Min2024}  method in the latter regime. The nonequilibrium dynamics  of QSL is initiated by coupling the Kitaev model~\cite{Kitaev2006, Matsuda2025}, as the widely studied example of QSL that is also exactly solvable in the closed case, to a bath of infinitely many bosonic modes described by the canonical Caldeira-Leggett model~\cite{Leggett1987}. From  $\hat{\rho}(t)$, we compute the time dependence of: GMN; quantum Fisher information (QFI)~\cite{Guehne2009a, Chiara2018,Shimokawa2025}, which is also experimentally accessible~\cite{Scheie2021, Hales2023}; spin-spin correlators; and the expectation value (EV) of the Wilson loop operator~\cite{Wilson1974, Hastings2005, Kitaev2006, Matsuda2025}. Such comprehensive dissection of properties of open QSL yields principal results summarized in Figs.~\ref{fig:Lindblad_results}--~\ref{fig:RC_results}. 

The paper is organized as follows. We first introduce key concepts and useful notation in Sec.~\ref{sec:methods}, which facilitates discussion of our results in Sec.~\ref{sec:results}. We also provide three appendices offering additional comparison of open Kitaev QLS vs. open quantum Heisenberg antiferromagnet on the same honeycomb lattice (Appendix~\ref{sec:appendixa}), analysis of finite size effects (Appendix~\ref{sec:appendixb}), and detailed derivation of effective steady-state Hamiltonian (Appendix~\ref{sec:appendixc}). We conclude in Sec.~\ref{sec:conclusions}.


\section{Models and methods}\label{sec:methods}

\subsection{Hamiltonian model}\label{sec:hamilton}

The Kitaev QSL Hamiltonian describes localized quantum spins $S=1/2$ that interact via highly anisotropic exchange interaction, as given by~\cite{Kitaev2006, Matsuda2025}
\begin{equation}\label{eq:Kitaev_Hamiltonian}
    \hat{H}_{\mathrm{QSL}} = -\sum_{\langle ij\rangle_x} J_x\hat{\sigma}_i^x \hat{\sigma}_j^x-\sum_{\langle ij\rangle_y} J_y\hat{\sigma}_i^y \hat{\sigma}_j^y-\sum_{\langle ij\rangle_z} J_z
    \hat{\sigma}_i^z \hat{\sigma}_j^z.
\end{equation}
Here, $(\hat{\sigma}_i^x, \hat{\sigma}_i^y, \hat{\sigma}_i^z)$ is the vector of the Pauli matrices describing spin at the site $i$ of the honeycomb lattice; $J_\mu$ is the magnitude of the Ising-like exchange interaction; and $\langle  ij\rangle_\mu$ denotes the nearest-neighbor (NN) bonds where \mbox{$\mu=x,y,z$}. This system is made open and dissipative by coupling it with one or many baths of infinitely many three-dimensional and isotropic bosonic modes~\cite{Leggett1987}, so that the Hamiltonian of the total system QSL+bath(s) becomes
$
    \hat{H}_\mathrm{tot}= \hat{H}_\mathrm{QSL}+\hat{H}_\mathrm{bath}+\hat{V}.
$
We distinguish two cases: ($i$) ``local coupling,'' where we couple each spin to an independent bath, so that
\begin{equation}\label{eq:local}
    \hat{H}_\mathrm{bath}=\sum_{ik} \omega_{ik} \hat{\mathbf{a}}^\dagger_{ik} \hat{\mathbf{a}}_{ik}, \: \hat{V}=\sum_{k} g_k \sum_{i\mu}\hat{\sigma}_i^\mu (\hat{a}^{\mu\dagger}_{ik}+\hat{a}^\mu_{ik});
\end{equation}
 ($ii$) ``global coupling,'' where only one bath is coupled to the total spin operator of QSL, so that
 \begin{equation}\label{eq:global}
     \hat{H}_\mathrm{bath}=\sum_{k} \omega_{k} \hat{\mathbf{a}}^\dagger_{k} \hat{\mathbf{a}}_{k}, \: \hat{V}=\sum_{k} g_k \sum_{i\mu} \hat{\sigma}_i^\mu (\hat{a}^{\mu\dagger}_{k}+\hat{a}^\mu_{k}).
 \end{equation}
 Here, bosonic operators \mbox{$\hat{\mathbf a}_{k} = (\hat a_{k}^x, \hat a_{k}^y, \hat a_{k}^z)^T$} and \mbox{$\hat{\mathbf a}_{ik} = (\hat a_{ik}^x, \hat a_{ik}^y, \hat a_{ik}^z)^T$} annihilate modes of frequency $\omega_{k}$ and $\omega_{ik}$, respectively. The spectral content of bosonic baths is specified~\cite{Leggett1987} by \mbox{$J(\omega) =2\pi \sum |g_k|^2\delta(\omega-\omega_k)$}. For our numerical calculations, we choose the Ohmic~\cite{Anders2022} bath with $J(\omega) =\Gamma \omega \exp(-\omega^2/2\Lambda^2)/[1-\exp(-\omega/T)]$, where $T$ is the temperature (we set $k_B=1$), $\Gamma$ is the single parameter characterizing the system-bath coupling, and $\Lambda$ sets the exponential cutoff for high frequencies~\cite{Rudner2020, Nathan2024}. Note that the hallmark of the Ohmic bath is $J(\omega) \propto\omega$ at low frequencies.

\subsection{Universal Lindblad quantum master equation}\label{sec:lindblad}

By assuming {\em small} $g_k$, a QME of the Lindblad type~\cite{Manzano2020, Lindblad1976} can be derived for the reduced density matrix $\hat{\rho}$ of QSL after bosonic bath(s) are traced over. However, standard approximations used in such derivations fail for systems with closely spaced (i.e., quasidegenerate) energy levels, as is the typical case with quantum magnets~\cite{GarciaGaitan2024, Szpak2024}. Thus, to obtain Lindblad QME for open QSL in the Markovian regime, we follow the procedure of Ref.~\cite{Rudner2020, Nathan2024}, yielding the so-called universal Lindblad QME. This QME~\cite{Rudner2020, Nathan2024} operates with three Lindblad operators $\hat{L}_{i \mu}$ per bath, so assuming $N$ such baths, we obtain the following QME
\begin{equation}
    \label{eq:Lindbladeq}
    d \hat{\rho}/dt = -i[\hat{H}_\mathrm{QSL},\hat{\rho}] +\sum_{i\mu}^N\left[ \hat{L}_{i\mu}\hat{\rho} \hat{L}_{i\mu}^\dagger -\frac{1}{2}\{ \hat{L}_{i\mu}^\dagger \hat{L}_{i\mu}, \hat{\rho}\}\right],
\end{equation}
where the Lamb-shift corrections to the Hamiltonian are neglected. We compute $\hat{L}_{i\mu}$ as a power series (where we use cutoff $N_L\leq 5$)
\begin{equation}
    \label{eq:expand}
    \hat{L}_{i\mu} = \sum_{n=0}^{N_L} c_n (\text{ad}_{\hat{H}_{\mathrm{QSL}}})^n [\hat{A}_{i\mu}], \: \: \: \: c_n=\frac{(-i)^n}{n!}\int_{-\infty}^\infty dt g(t) t^n,
\end{equation}
thereby evading exact diagonalization of the full QSL Hamiltonian~\cite{Rudner2020,Schnell2025}. Here \mbox{$\text{ad}_{\hat{H}} [\hat{X}] = [\hat{H},\hat{X}]$};  \mbox{$\hat{A}_{i\mu}=\hat{\sigma}_i^\mu$} in the ``local coupling'' case [Eq.~\eqref{eq:local}] or \mbox{$\hat{A}_{\mu} =\sum_j\hat{\sigma}_j^\mu$} in the ``global coupling'' case  [Eq.~\eqref{eq:global}]; and the jump correlator function is defined via the Fourier transform of the spectral function of the bath,
\mbox{$g(t)=\frac{1}{\sqrt{2\pi}}\int_{-\infty}^{\infty} d\omega\sqrt{J(\omega)} e^{-i \omega t}$}. The universal Lindblad QME is solved using the quantum trajectories method~\cite{Daley2014}, with up to $\sim 1000$ trajectories, as implemented in the  \texttt{QuTiP} package~\cite{Johansson2013a, Lambert2024}. In these calculations, we use $\Gamma=0.001 J_z$ and $\Lambda=50T$.

\subsection{Tensor network methodology for open quantum systems}\label{sec:tn}

For strong coupling of a system to a structured environment, responsible for non-perturbative effects~\cite{ReyesOsorio2026} and non-Markovian dynamics~\cite{Breuer2007, Vega2017}, there is currently no universal approach akin to the Lindblad QME for the Markovian regime. Among a handful~\cite{Fux2023, Fux2024, Cygorek2024a, ReyesOsorio2026,Sun2024a, Xu2024,Lindoy2025} of very recently developed methods that can handle many quantum spins as the system and arbitrary spectral function or temperature of dissipative bosonic environment, we chose the TN methodology~\cite{Fux2023, Fux2024, Cygorek2024} implemented as process tensor matrix product operator (PT-MPO). The PT-MPO combined with time-evolving block decimation (TEBD), as available in the \texttt{OQuPy} package~\cite{Fux2024}, allows us to perform real-time evolution of open QSL. Such evolution is restricted to coupling the same local bosonic bath to a single component of spin at site $i$, thereby making it a particular case of the ``local coupling'' in Eq.~\eqref{eq:local}. In PT-MPO+TEBD calculations, we use $\Lambda = 4 J_z$ and stronger $\Gamma =0.1 J_z$. For the system hosting only the NN interactions between spins, we considered supersites composed of two physical spins so that the hexagonal ladder becomes effectively a chain. The PT-MPO is then created with a maximum tolerance of $0.01$ and the time evolution is performed, keeping a maximum relative error of $\mathcal{O}(10^{-5})$~\cite{Fux2024}. Since PT-MPO+TEBD calculations have difficulties starting from an entangled initial state, we use unentangled \mbox{$\hat \rho(t=0) = (\mathbf{I} \otimes \mathbf{I} \otimes \dots \otimes \mathbf{I})/2^N$} for $N$ spins, whose simplicity is also computationally favorable. Here the Kronecker product contains $N/2$ terms $\mathbf{I}$, where $\mathbf{I}$ is the unit $4\times4$ matrix, so that $\mathbf{I}/4$ is the density matrix of two spins within the supersite. This choice does not affect conclusions in the long-time limit, as entanglement is quickly built up dynamically [Figs.~\ref{fig:TEMPO_results}(a) and \ref{fig:TEMPO_results}(b)] despite zero entanglement in $\hat \rho(t=0)$.

\subsection{Reaction coordinate + polaron methodology}

Since time-dependent TN methods easily encounter an ``entanglement barrier''~\cite{Lerose2023,Rams2020,Foligno2023}, which  prevents reaching truly long evolution times, to access the long-time limit of non-Markovian dynamics, we also employ the RC  method~\cite{Schaller2014} combined with polaron transformation~\cite{Anto-Sztrikacs2023}. This approach produces an effective Hamiltonian that can accurately~\cite{Brenes2024} describe the steady-state of non-Markovian dynamics. The same strategy has been applied previously to classify possible magnetic ordering of different steady-states reached by the dissipative dynamics of quantum spin chains~\cite{Min2024}. The RC method~\cite{Schaller2014}, which is based on the Bogoliubov transformation, introduces a new RC bosonic mode per each bath. They are created (annihilated) by an operator $\hat{\mathbf{b}}_i^\dagger(\hat{\mathbf{b}}_i)$, which is strongly coupled to the system but weakly coupled to the remaining modes of the bath. This transforms the Hamiltonian $\hat{H}_\mathrm{tot}$ into
\begin{widetext}
\begin{align}
    \label{eq:RC_Hamiltonian}
    \hat{H}_\mathrm{RC} = \hat{H}_\mathrm{QSL} + \lambda \sum_i \hat{A}_{i\mu}(\hat{b}_i^\mu+\hat{b}_i^{\mu\dagger}) +\Omega \sum_i\hat{\mathbf{b}} _i^\dagger \hat{\mathbf{b}}_i +\hat{H}_\mathrm{RC-B}+\hat{\tilde{H}}_\mathrm{bath},
\end{align}
\end{widetext}
where $\lambda$ is the strength of the coupling between the RC mode and the system; $\Omega$ is the frequency of the RC mode; \mbox{$\hat{H}_\mathrm{RC-B}=\sum_i\sum_{k>1,\mu}\tilde{g_k}(\hat{b}_i^\mu+\hat{b}_i^{\mu\dagger})(\hat{c}_{ik}^\mu+\hat{c}_{ik}^{\mu\dagger})$} describes the coupling between the RC mode and the remaining bosonic bath; and $\hat{\tilde{H}}_\mathrm{bath}=\sum_{i,k>1}\tilde{\omega}_k \hat{\mathbf{c}}^\dagger_{ik}\hat{\mathbf{c}}_{ik}$ is the Hamiltonian of residual bosonic modes. Here bosonic operators \mbox{$\hat{\mathbf c}_{ik} = (\hat c_{ik}^x, \hat c_{ik}^y, \hat c_{ik}^z)^T$}, obtained by the Bogoliubov transformation of the original operators in $\hat{H}_\mathrm{tot}$, have frequency $\tilde{\omega}_k$.
The second step of the RC + polaron methodology involves a polaron transformation~\cite{Anto-Sztrikacs2023, Min2024} per bath, incorporating the RC-system interaction strength directly into the system Hamiltonian. This transforms Hamiltonian in Eq.~\eqref{eq:RC_Hamiltonian} into
    $\hat{H}_\mathrm{RC-P} = \prod_{i\mu} \hat{U}_{P}^{i\mu} \hat{H}_\mathrm{RC}\hat{U}_P^{i\mu\dagger},
$
where \mbox{$\hat{U}_P^{i}=\prod_\mu\exp{(\lambda/\Omega (\hat{b}_i^{\mu\dagger}-\hat{b}_i^\mu)\hat{A}_{i\mu}})$} is the polaron transformation associated with the $i$-th bath. The final step projects the Hamiltonian onto the GS of each RC, which is an approximation valid in the low-temperature limit~\cite{Anto-Sztrikacs2023}. The resulting effective Hamiltonian is then given by
\begin{equation}
    \label{eq:effective_Hamiltonian}
    \hat{H}_\mathrm{eff} = \mathrm{Tr}_{\mathrm{RC}}(\hat{\Pi}_0 \hat{H}_\mathrm{RC-P}\hat{\Pi}_0),
\end{equation}
where $\hat{\Pi}_0$ is the projection operator onto the product of GSs of each RC variable. Note that special care has to be taken in the case where the set of coupling operators $\{ \hat{A}_{i\mu}\}$ do not commute~\cite{Garwola2024}, as is our case. 

Application of Eq.~\eqref{eq:effective_Hamiltonian} to $\hat{H}_\mathrm{tot}$ of open Kitaev QSL yields, for the ``local coupling'' case [Eq.~\eqref{eq:local}], the following effective Hamiltonian
\begin{widetext}
\begin{align}
    \label{eq:Heff_local}
    \hat{H}_{\mathrm{eff}}^\mathrm{loc} = -\kappa_{J_K}^\mathrm{loc}(\lambda/\Omega) \left[\sum_{\langle ij\rangle_x} J_x  \hat{\sigma}_i^x \hat{\sigma}_j^x +\sum_{\langle ij\rangle_y} J_y  \hat{\sigma}_i^y \hat{\sigma}_j^y+\sum_{\langle ij\rangle_z} J_z  \hat{\sigma}_i^z \hat{\sigma}_j^z \right].
\end{align}
\end{widetext}
We see that this Hamiltonian still has the Kitaev form in Eq.~\eqref{eq:Kitaev_Hamiltonian}, but its exchange interactions are renormalized by prefactor $\kappa_{J_K} (\lambda/\Omega)$  whose dependence on QSL-bath coupling is plotted in Fig.~\ref{fig:RC_results}(a). Notably, as we couple each spin to three independent baths via non-commuting operators, the strength of the Kitaev exchange interaction does not decay to zero but instead stabilizes at a finite value [Fig.~\ref{fig:RC_results}(a)].

In the ``global coupling'' case [Eq.~\eqref{eq:global}], an additional ferromagnetic Heisenberg exchange interaction term between spins on NN sites, as well as {\em all-to-all} ferromagnetic Heisenberg exchange, emerge due to the bath:
\begin{widetext}
\begin{align}
    \label{eq:Heff_global}
    \hat{H}_\mathrm{eff}^\mathrm{glob} = -\sum_{\langle ij\rangle_\mu} J_\mu \kappa^\mathrm{glob}_{J_K}(\lambda/\Omega) \hat{\sigma}^\mu_i \hat{\sigma}_j^\mu -\sum_{\langle ij\rangle}\sum_\mu J_\mu \kappa^\mathrm{glob}_{J_H}(\lambda/\Omega) \hat{\boldsymbol{\sigma}}_i \cdot \hat{\boldsymbol{\sigma}}_j 
    - \sum_{ij\mu}\frac{\lambda^2}{\Omega} \left(\kappa_{J_K}^\mathrm{glob}(\lambda/\Omega)\hat{\sigma}^\mu_i\hat{\sigma}^\mu_j + \kappa_{J_H}^\mathrm{glob}(\lambda/\Omega) \hat{\boldsymbol{\sigma}}_i \cdot \hat{\boldsymbol{\sigma}}_j\right).
\end{align}
\end{widetext}
Here $\kappa_{J_K}^\mathrm{glob}(\lambda/\Omega)$ and $\kappa_{J_H}^\mathrm{glob}(\lambda/\Omega)$ are the prefactors renormalizing Kitaev and Heisenberg exchange interactions, respectively. They are plotted in Fig.~\ref{fig:RC_results}(b) as a function of QSL-bath coupling. Note that detailed derivation of Eqs.~\eqref{eq:Heff_local} and ~\eqref{eq:Heff_global} is presented in appendix~\ref{sec:appendixc}.

\section{Results and discussion}\label{sec:results}

\begin{figure}[t!]
    \centering
    \includegraphics[width=8.5cm]
    {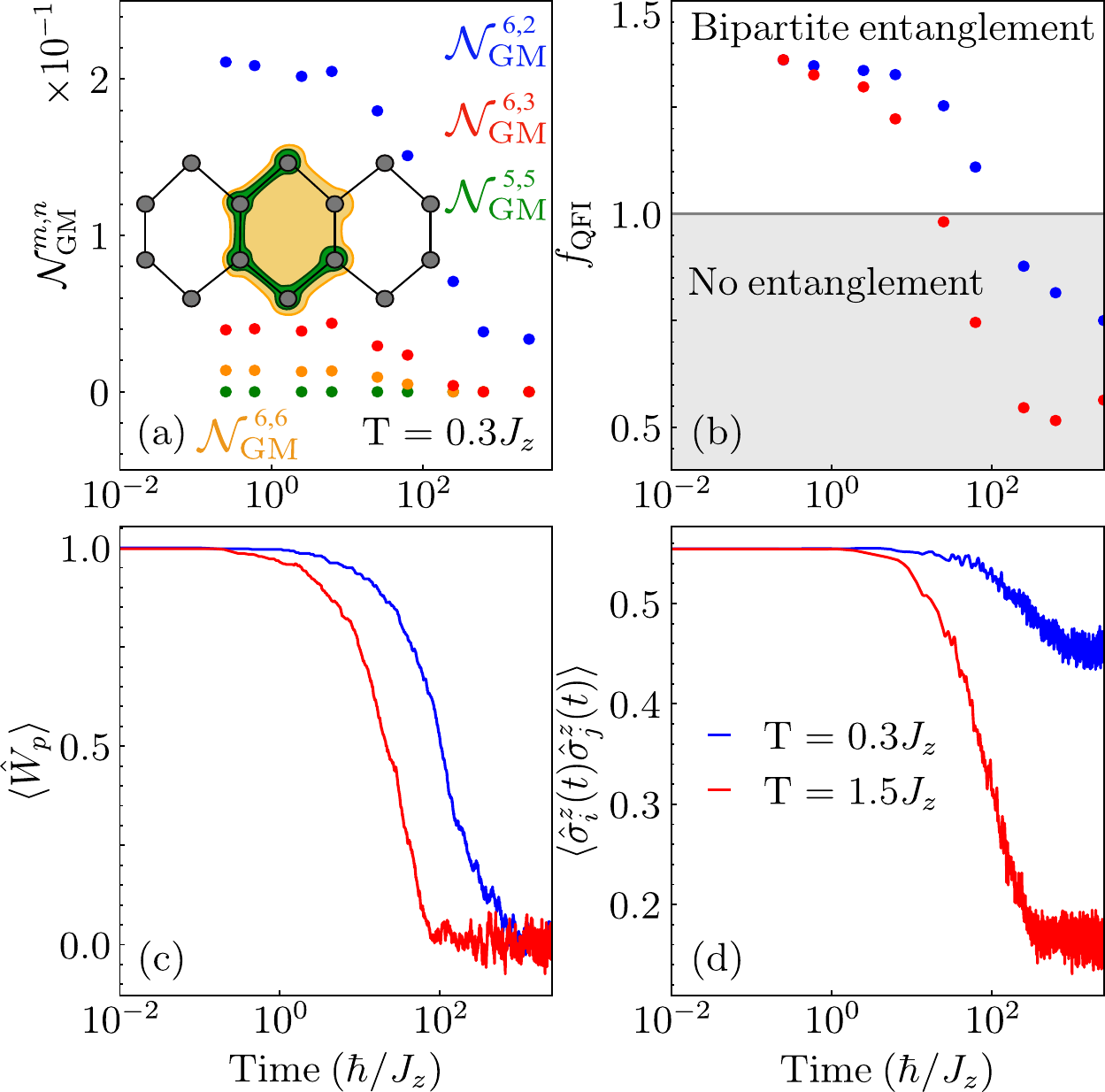}
    \caption{Time dependence in the {\em Markovian} regime, as computed by the universal Lindblad QME~\cite{Rudner2020, Nathan2024}, of: (a) $n$-party GMN~\cite{Lyu2025,Jungnitsch2011,Hofmann2014} $\mathcal{N}_\mathrm{GM}^{m,n}$ for two different loopy (yellow) and non-loopy (green) subregions enclosed within the colored middle hexagon and each party composed of a single spin, here $m$ indicates the number of spins in each subregion; (b) QFI [Eq.~\eqref{eq:QFI}] for the wavevector $\mathbf{k}=(0,0)$; (c) Wilson loop operator $\langle \hat{W}_p\rangle(t)$; and (d) equal-time spin-spin correlator $\langle \hat{\sigma}_i^z(t) \hat{\sigma}_{j}^z(t) \rangle$ for two NN sites $i$ and $j$. The system considered is the gapless Kitaev QSL [Eq.~\eqref{eq:Kitaev_Hamiltonian}] composed of $N=14$ spins locally coupled to many baths [Eq.~\eqref{eq:local}]. The temperature of bosonic baths for (b)--(d) is indicated in panel (d), while in (a) it is set to $T=0.3J_z$.}
    \label{fig:Lindblad_results}
\end{figure}

\begin{figure}[t!]
    \centering
    \includegraphics[width=8.5cm]
    {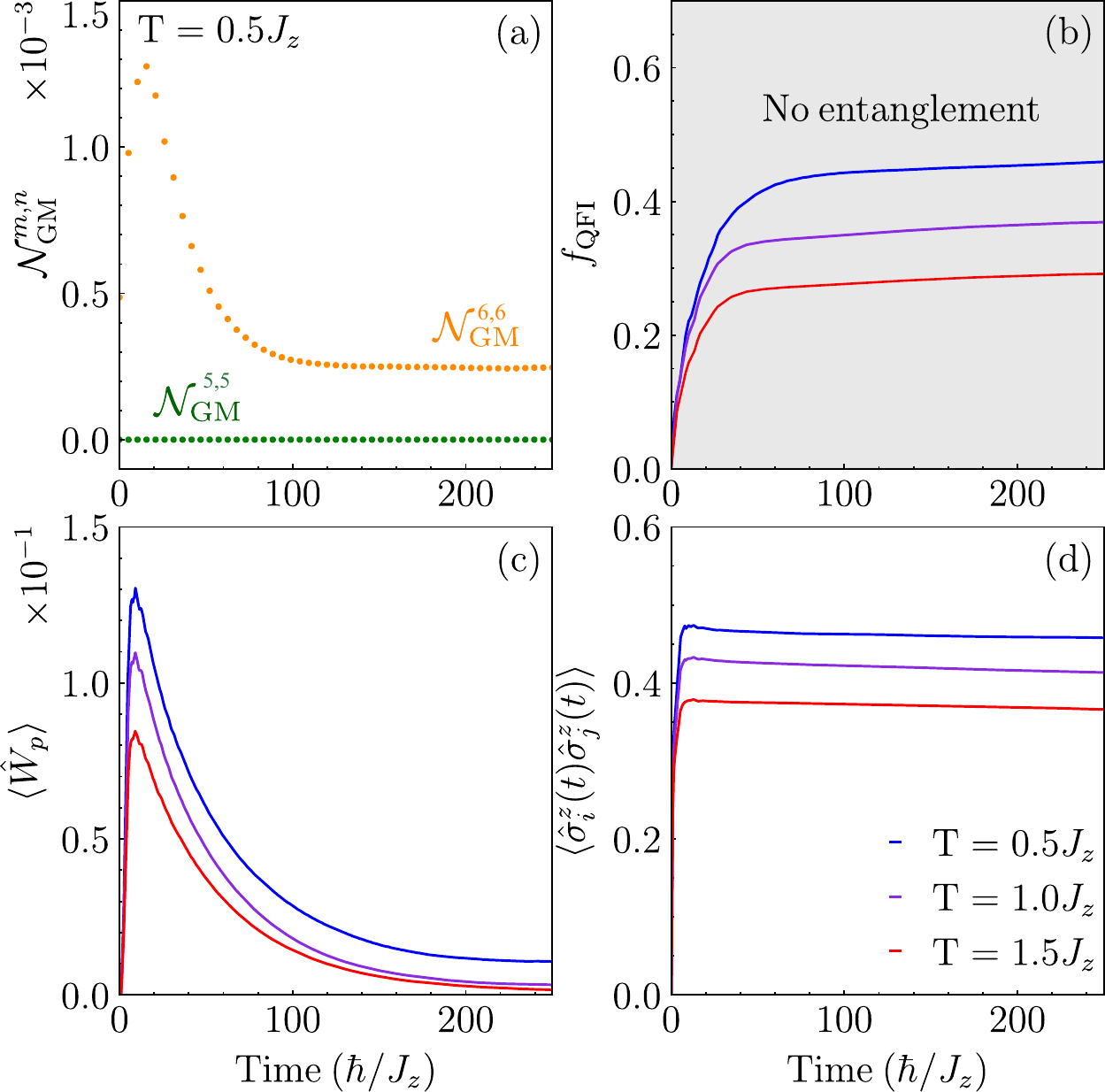}
    \caption{The same information as in Fig.~\ref{fig:Lindblad_results}, but for the {\em non-Markovian} regime, as computed via PT-MPO+TEBD methodology~\cite{Fux2023, Fux2024, Cygorek2024, Cygorek2024a}. The temperature of bosonic baths for (b)--(d) is indicated in panel (d), while in (a) it is set to $T=0.5J_z$. Note that orange dots in panel (a) start from zero due to the unentangled initial state required for PT-MPO+TEBD calculations, unlike orange dots in Fig.~\ref{fig:Lindblad_results}(a), where the initial state is a highly entangled GS of QSL.}
    \label{fig:TEMPO_results}
\end{figure}

We consider the Kitaev QSL composed of $N=14$ spins $S=1/2$ located on the sites of  three hexagons, for which $J_z=J_x=J_y=1$ sets the unit of energy. Since evaluation of GMN is computationally demanding (as it corresponds to a constrained semidefinite program~\cite{Lyu2025, Sabharwal2025}),  it is currently restricted to a few spins~\cite{Lyu2025,Jungnitsch2011}. Therefore, we focus on the loopy subregion of 6 spins [residing on sites within the orange hexagon in the inset of Fig.~\ref{fig:Lindblad_results}(a)] or the non-loopy one of 5 spins [encompassed by the green bonds in the inset of Fig.~\ref{fig:Lindblad_results}(a)]. For these subregions, we compute GMN $\mathcal{N}_\mathrm{GM}^{m,n}$ for $n$ parties within a subregion of $m=6$ or $5$ spins. Here, each party consists of $m/n$ spins (i.e., we group spins into composite subsystems before evaluating multipartite entanglement). Time evolution of $\mathcal{N}_\mathrm{GM}^{6,n}(t)$ and $\mathcal{N}_\mathrm{GM}^{5,n}(t)$ is shown in Figs.~\ref{fig:Lindblad_results}(a) and \ref{fig:TEMPO_results}(a) for the Markovian and non-Markovian regimes, respectively. For reference, the same subregions were analyzed in the closed Kitaev QSL in Ref.~\cite{Lyu2025}, whose density matrix is defined by $\hat \rho_\mathrm{subregion}=\mathrm{Tr}_{\mathrm{other}}|\mathrm{GS} \rangle \langle\mathrm{GS}|$. In contrast, we use $\hat \rho_\mathrm{subregion}(t)=\mathrm{Tr}_{\mathrm{other}} \hat \rho(t)$ in our calculations for open Kitaev QSL. A finite GMN indicates that for {\em all possible}  bipartitions into sub-subregions 1 and 2 of a chosen subregion of $m$ parties of QSL, its density matrix remains entangled, $\hat\rho_\mathrm{subregion} \neq \sum_n p_n \hat\rho_n^{(1)} \otimes \hat\rho_n^{(2)}$. Thus, diagnostics of entanglement offered by GMN is far more general than either the von Neumann and Renyi entropies~\cite{Brydges2019}, which apply only to pure quantum states; or logarithmic negativity routinely employed~\cite{Peres1996, Lu2023, Fan2024, Suresh2024} to detect entanglement in mixed quantum states but only by considering a single bipartition. For example, the number of bipartitions examined by GMN for the loopy hexagonal subregion of six parties in Figs.~\ref{fig:Lindblad_results}(a) and ~\ref{fig:TEMPO_results}(a) is $2^6-1$. In the Markovian regime, non-loopy subregions exhibit zero GMN [green dots in Fig.~\ref{fig:Lindblad_results}(a)] at all times, as is the case of closed Kitaev QSL~\cite{Lyu2025}; while GMN of loopy subregions decays with time [orange, red, and blue dots in Fig.~\ref{fig:Lindblad_results}(a)] and saturates to a finite steady value only for $n=2$. For the closed Kitaev QSL, it has been pointed out~\cite{Lyu2025} that genuine QSL features are signaled by a non-vanishing $\mathcal{N}_\mathrm{GM}^{6,6}$ value. Thus, the finite steady-state value of $\mathcal{N}_\mathrm{GM}^{6,2}(t)$ indicates an intermediate regime where QSL remains at least bipartite entangled while losing its finer entanglement structure signaled by a finite $\mathcal{N}_\mathrm{GM}^{6,6}(t)$ value~\footnote{The same intermediate state, with $\mathcal{N}^{6,6}_\mathrm{GM}=0$, has been observed in very recent calculations of GMN for closed Kitaev QSL~\cite{Lyu2025} in an interval of external magnetic field, as well as for Kitaev QSL at finite temperature~\cite{Sabharwal2025}.  We estimate this state to exist for temperatures $ \in [0.3J_z,1J_z]$ in the case of the Markovian regime, but precise boundaries of this interval would require examining larger system sizes.}. This is reminiscent of the conclusions of finite-temperature calculations focused on fractionalized excitations of Kitaev QSL~\cite{Nasu2015} where a crossover behavior was found between two peaks of specific heat versus temperature. Surprisingly, despite dissipation, in the non-Markovian regime, the finer GMN of loopy subregions ($\mathcal{N}_\mathrm{GM}^{6,6}(t)$) remains finite at all times [orange dots in Fig.~\ref{fig:TEMPO_results}(a)] up to a much higher temperature $T\leq0.7J_z$ than in the case of Markovian dynamics. This indicates that elusive QSL features can be harvested in a higher temperature regime by coupling to a more structured dissipation structure.

It is also worth comparing the time evolution of GMN in open QSL vs. open quantum antiferromagnet (QAF), both of which are defined on the same honeycomb lattice. As shown in Fig.~\ref{fig:figS1} of Appendix~\ref{sec:appendixa}, GMN remains nonzero in {\em both} non-loopy [in contrast to zero value for open QSL in Fig.~\ref{fig:Lindblad_results}(a)] and loopy subregions in the course of Markovian dynamics.


\begin{figure}[t!]
    \centering
    \includegraphics[width=8.5cm]
    {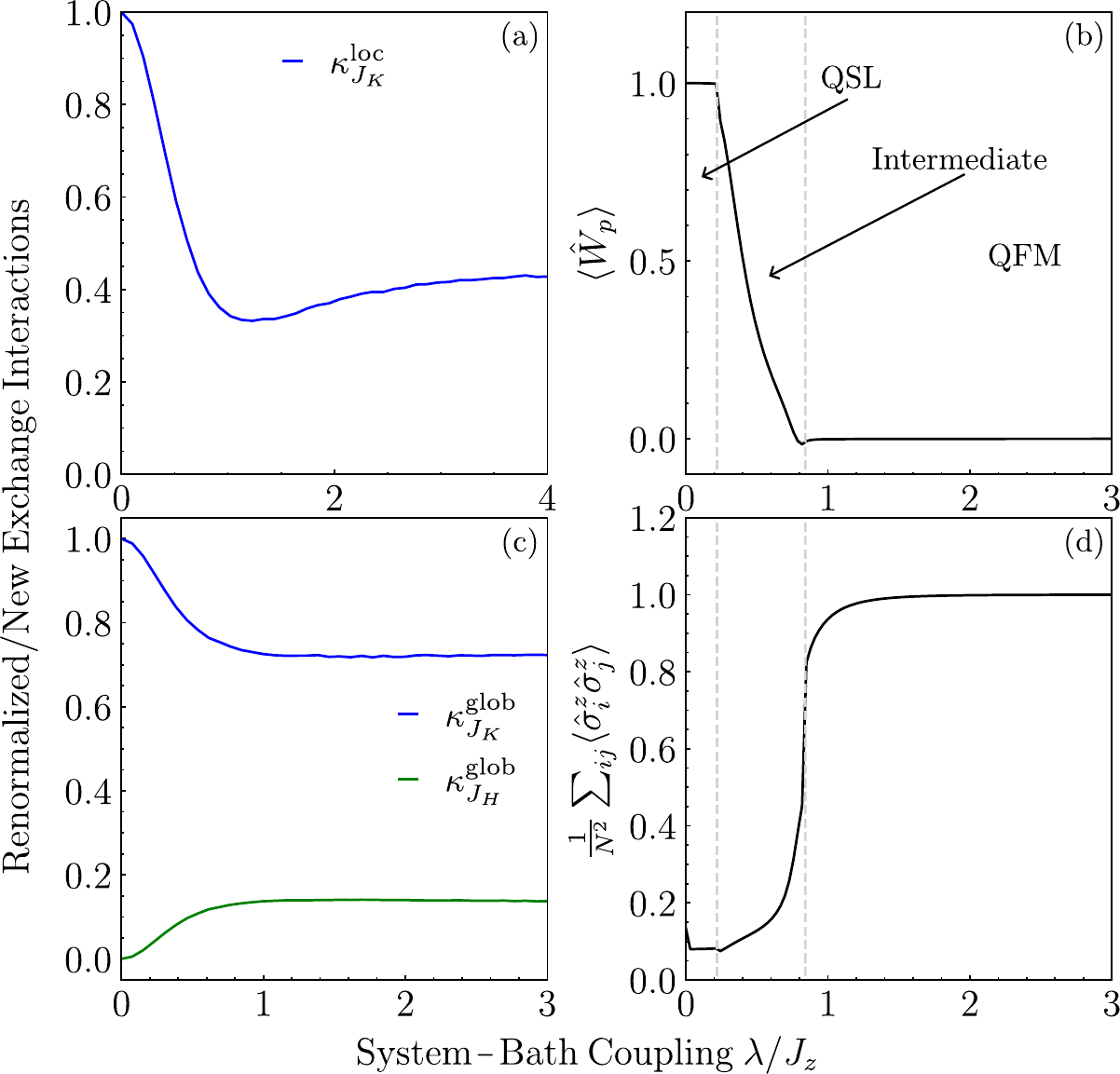}
    \caption{Renormalized exchange interactions of the Kitaev QSL Hamiltonian [Eq.~\eqref{eq:Kitaev_Hamiltonian}] for:  (a) ``local coupling'' to many baths [Eq.~\eqref{eq:local}]; and (c) ``global coupling'' to a single bath [Eq.~\eqref{eq:global}]. (b) EV of Wilson loop operator for the case of ``global coupling.'' (d) Static FM structure factor corresponding to panel (c). The results in (b) and (d) were obtained by diagonalizing the effective Hamiltonian [Eq.~\eqref{eq:Heff_global}] produced by the RC + polaron methodology, so they describe steady state in the long-time limit of {\em non-Markovian} dynamics generated by arbitrary strong coupling to a single global bath. The frequency of the RC was set to $\Omega=8J_z$. The $\lambda/J_z \rightarrow 0$ limit corresponds to the Markovian dynamics of Fig.~\ref{fig:Lindblad_results}.}
    \label{fig:RC_results}
\end{figure}

While the entanglement entropy of pure states of cold atom simulators of quantum magnets has been directly measured~\cite{Brydges2019}, this is not possible in the case of solid-state magnetic materials. Instead, recent efforts have focused on extracting entanglement witnesses~\cite{Friis2018, Guehne2009a} from experiments, as functionals of the density matrix not requiring its full tomography~\cite{White1999}, for solids both in~\cite{Scheie2021, Laurell2024} and out~\cite{Hales2023, Ren2024} of equilibrium. One such witness is QFI~\cite{Scheie2021, Laurell2024, Hales2023, Ren2024,Shimokawa2025}, whose recent study in QSL~\cite{Shimokawa2025} motivates its computation in Figs.~\ref{fig:Lindblad_results}(b) and \ref{fig:TEMPO_results}(b) using the following expression~\cite{Toth2012, Hyllus2012, Pezze2009}
\begin{align}
    f_\mathrm{QFI}(\mathbf{k}, t) &= 2 \sum_{\ell, \ell^\prime} \frac{(\lambda_\ell(t)-\lambda_{\ell^\prime}(t))^2}{N[\lambda_\ell(t)+\lambda_{\ell^\prime}(t)]} |\langle\ell(t) | \hat{\mathcal{A}}(\mathbf{k})|\ell^\prime(t)\rangle|^2,
    \label{eq:QFI}
\end{align}
where $\lambda_\ell(t)$ and $|\ell(t)\rangle$ denote eigenvalues and eigenvectors, respectively. Additionally, the sum runs over all eigenvalues such that \mbox{$\lambda_\ell(t)+\lambda_{\ell^\prime}(t)>0$}, $\hat{\mathcal{A}}(\mathbf{k})=\sum_i e^{i \mathbf{k}\cdot\mathbf{r}_i} \hat{\sigma}_i^z/2$ is a proper collective many-body operator with $\mathbf{r}_i$ being the position of the site $i$ of the lattice, and $\mathbf{k}$ is the crystalline wavevector. To facilitate direct comparison with GMN results on Figs.~\ref{fig:Lindblad_results}(a) and~\ref{fig:TEMPO_results}(a), QFI is computed on the same hexagonal subregion. The time-dependent QFI~\cite{Hales2023, Ren2024, Suresh2024} can witness $n$-partite entanglement by exhibiting $f_\mathrm{QFI}(\mathbf{k},t)\leq 4S^2 n$. Thus, $f_\mathrm{QFI}(\mathbf{k},t)>1$ indicates at least a bipartite entangled mixed state of the whole system. Figures~\ref{fig:Lindblad_results}(b) and \ref{fig:TEMPO_results}(b) show that open dynamics eventually render QFI entanglement detection criteria unusable as it unavoidably trespasses the critical $f_\mathrm{QFI}(t)=1$ threshold. Comparison of time-dependent QFI in the Markovian and non-Markovian regimes [e.g., Fig.~\ref{fig:TEMPO_results}(a) vs. Fig.~\ref{fig:TEMPO_results}(b)] shows how intricate entanglement formation signaled by GMN on a loopy subregion can be {\em completely missed} by the QFI entanglement criterion.

To complement understanding of observed entanglement dynamics, we  analyze EV of operators associated with fractionalized excitations. For Kitaev QSL, one often employs~\cite{Nasu2015}  Wilson operator, \mbox{$\hat{W}_p = \hat \sigma_1^z \hat \sigma_2^y \hat \sigma_3^x \hat \sigma_4^z \hat \sigma_5^y \hat \sigma_6^x$}, for a loop of sites, such as 1 to 6 forming the first hexagon in the inset of Fig.~\ref{fig:Lindblad_results}(a) that we choose. Its EV  for a closed Kitaev QSL at zero temperature, $\langle \hat{W}_p\rangle =1$,  indicates ordering of $\mathbb{Z}_2$ fluxes as one of the fractionalized excitations within the model. Thus, it has  been previously employed to study how they disorder due to thermal fluctuations~\cite{Nasu2015}. For open Kitaev QSL, we find that $ \langle\hat W_p \rangle(t)$ decays with time in Figs.~\ref{fig:Lindblad_results}(c) and \ref{fig:TEMPO_results}(c), with faster decay at higher temperatures. However, it can remain nonzero at times at which GMN vanishes in the Markovian regime [Fig.~\ref{fig:Lindblad_results}(c)], thereby again suggesting that GMN is a superior indicator of the limits of the persistence of the QSL phase~\footnote{The same conclusion about GMN being a superior indicator of equilibrium QSL at finite temperature was obtained in Ref.~\cite{Sabharwal2025}, where the EV of the Wilson operator remains nonzero at temperatures at which GMN vanishes.}. In the non-Markovian regime, both GNN and $ \langle\hat W_p \rangle(t)$ plateau in the long time limit, on the proviso that the bosonic bath temperature is sufficiently low. 

We also examine the time evolution of the equal-time spin-spin correlator $\langle \hat{\sigma}_i^z(t) \hat{\sigma}_j^z(t) \rangle$ for two NN sites $i$ and $j$ in open Kitaev QSL. The same quantity in {\em closed} Kitaev QSL~\cite{Baskaran2007, Knolle2015, Takegami2025} is {\em zero} beyond NN sites, thereby signifying a lack of long-range magnetic ordering. We find the same feature in open QSL [Figs.~\ref{fig:Lindblad_results}(d) and \ref{fig:TEMPO_results}(d)] when ``local coupling'' [Eq.~\eqref{eq:local}] to many baths is used. This correlator then decays with increasing time or temperature, with faster decay [compare Fig.~\ref{fig:Lindblad_results}(d) vs. Fig.~\ref{fig:TEMPO_results}(d)] in the Markovian regime
where its decay is also highly correlated with the decay of QFI. 

Finally, to examine the case of a {\em single} global bosonic bath, we employ the RC + polaron method. This method also makes it possible to construct an effective Hamiltonian~\footnotemark[1] whose low-energy eigenstates provide insight into the long-time limit properties of non-Markovian dynamics at arbitrary strength of QSL-bath coupling. For the ``local coupling'' case, we observe that the effective Hamiltonian corresponds to a renormalization of the original Kitaev model [Eq.~\eqref{eq:Heff_local}], thereby explaining why the results of Figs.~\ref{fig:Lindblad_results} and~\ref{fig:TEMPO_results} preserve characteristic features (such as spin-spin correlations being nonzero only for nearest neighbors) discussed in prior literature~\cite{Baskaran2007, Knolle2015, Takegami2025} on closed Kitaev QSLs. Unlike the ``local coupling'' case [Eq.~\eqref{eq:local}] studied in Figs.~\ref{fig:Lindblad_results} and~\ref{fig:TEMPO_results}, in the ``global coupling'' case [Eq.~\eqref{eq:global}] we find new [green curve in Fig.~\ref{fig:RC_results}(c)] bath-induced exchange interactions between spins, as well as renormalization of the old ones [blue curves in Figs.~\ref{fig:RC_results}(a) and~\ref{fig:RC_results}(c)]. These new interactions include all-to-all ferromagnetic Heisenberg exchange, $\lambda^2/\Omega \kappa_{J_H}^\mathrm{glob} (\lambda/\Omega)$ [Eq.~\eqref{eq:Heff_global}], which grows with the strength of the system-bath coupling $\lambda$. Therefore, a crossover from QSL to a quantum ferromagnet (QFM) could be expected [Fig.~\ref{fig:RC_results}(d)], as confirmed by computing the static ferromagnetic structure factor [Fig.~\ref{fig:RC_results}(d)] or the EV of the Wilson loop operator [Fig.~\ref{fig:RC_results}(b)]. The same new exchange interactions then dramatically modify the well-known result~\cite{Baskaran2007}, $\langle \hat{\sigma}_i^\mu \hat{\sigma}_j^\nu\rangle \propto \delta^{\mu \nu} \delta_{\langle ij \rangle_\mu}$, for closed Kitaev QSL as we find such a spin-spin correlator to extend beyond NN sites [Fig.~\ref{fig:RC_results}(d)].

\section{Conclusions}\label{sec:conclusions}

In contrast to long-term studied bipartite~\cite{Grover2013, Kitaev2006a, Levin2006, Jiang2012a, Wildeboer2017, Depenbrock2012, Zou2016}, or very recently initiated  multipartite~\cite{Lyu2025}, entanglement of {\em closed} QSLs, its fate in QSLs made {\em open} by coupling them to a dissipative structured environment remains largely unexplored despite its relevance for experiments and quantum computing applications. Our real-time evolution of open Kitaev QSL, via Lindblad QME in the Markovian regime or TN methods in the non-Markovian regime, demonstrates how the {\em multipartite} entanglement quantified by GMN can be remarkably robust in the latter regime. We also unravel how non-vanishing GMN is accompanied by nonzero values of other quantities like QFI, EV of the Wilson loop operator, and spin-spin correlator. However, the fact that the latter three quantities can be nonzero even when GMN becomes zero at the chosen time and temperature signals an intermediate entangled regime~\cite{Lyu2025, Sabharwal2025, Nasu2015} that is not entirely QSL anymore. Taken together, our results reveal a clear hierarchy of entanglement features in open Kitaev QSLs: multipartite entanglement detected by the GMN is the most fragile, as it is lost before $\mathbb{Z}_2$ fluxes disorder or QFI  and spin-spin correlators vanish. Furthermore, non-Markovian dynamics can induce novel effects, beyond the na\"{i}ve expectation of the environment simply diminishing the entanglement, such as nonzero GMN at higher temperatures or additional exchange interactions between spins. Thus, our findings open pathways for engineering  dissipation~\cite{Harrington2022} and decoherence~\cite{Sohal2025,Ellison2025,Wang2025}  to tailor properties of QSL candidate materials for topological quantum computing, which we relegate to future studies. Finally, Fig.~\ref{fig:figS2} of Appendix~\ref{sec:appendixb} demonstrates that finite-size effects do not change any of our 
major conclusions. 

This work was supported by the U.S. National Science Foundation (NSF) under Grant No. DMR-2500816. The supercomputing time was provided by DARWIN (Delaware Advanced Research Workforce and Innovation Network), which is supported by NSF Grant No. MRI-1919839.

\appendix

\begin{figure}
    \centering
    \includegraphics[width=8.5cm]{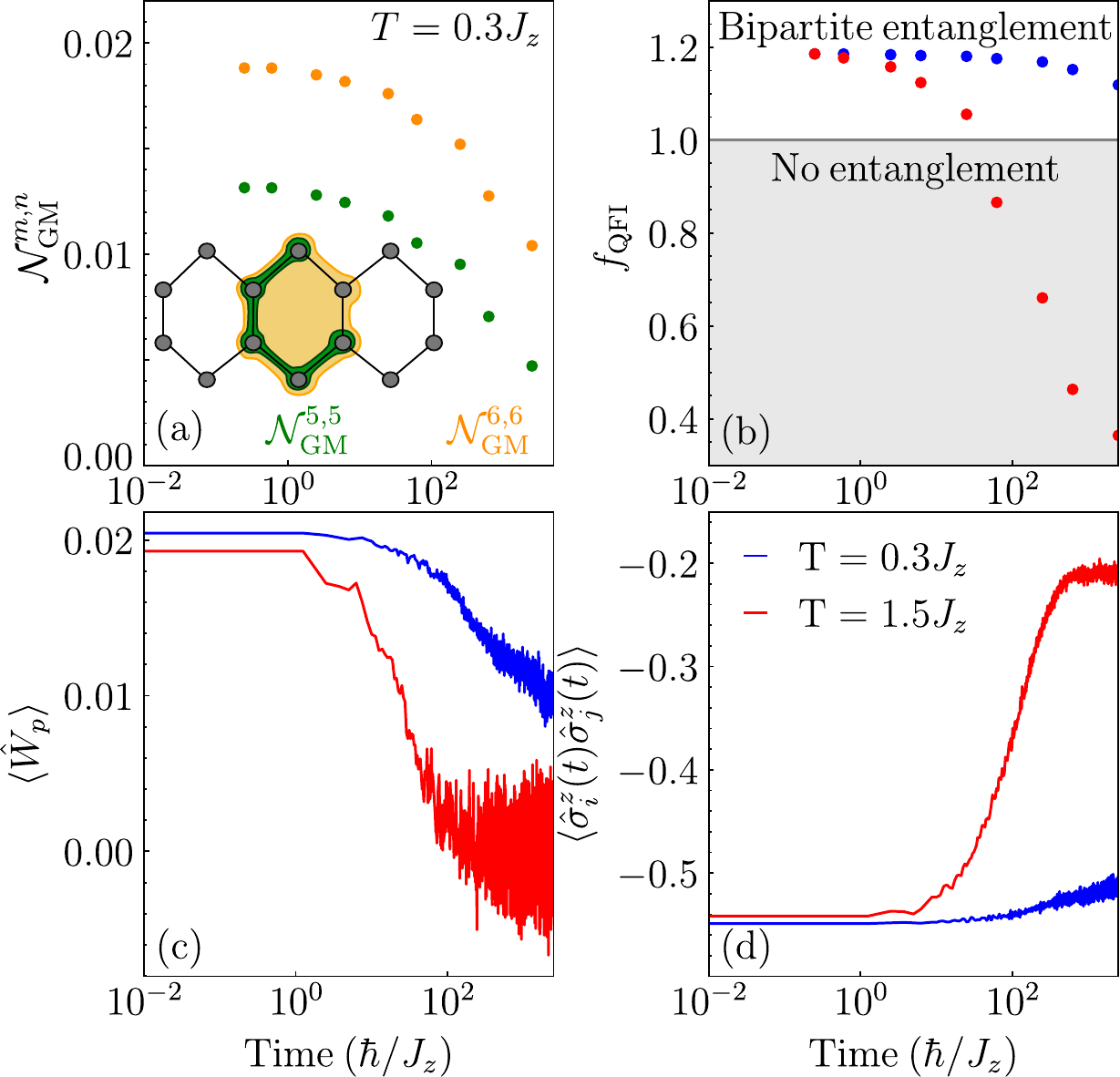}
    \caption{The same information as in Fig.~\ref{fig:Lindblad_results}, but for {\em quantum antiferromagnet} on the honeycomb lattice [inset of panel (a)] in the {\em Markovian} regime. The temperature of bosonic baths for (b)--(d) is indicated in panel (c), while in (a) is set to $T=0.3 J$. Note that the wavevector at which quantum Fisher information is computed in panel (b) is $\mathbf{k}=(\pi,\pi)$.}
    \label{fig:figS1}
\end{figure}

\section{Time-evolution of open quantum Heisenberg antiferromagnet on the honeycomb lattice in the Markovian regime}\label{sec:appendixa}

Figure~\ref{fig:figS1} gives the same information as  Fig.~\ref{fig:Lindblad_results}, but for the   quantum antiferromagnetic (QAF) Heisenberg model defined on the honeycomb lattice and time-evolved using the universal Lindblad quantum master equation (QME)~\cite{Rudner2020,Nathan2024}. Note that the complexity of such honeycomb QAF, despite lacking frustration akin to QSL, has motivated very recent neutron scattering  experiments~\cite{Hernandez2025} on its realization via two-dimensional magnetic materials. The Hamiltonian of quantum $S=1/2$ antiferromagnet is given by
\begin{equation}
    \label{eq:Heisenberg_Hamiltonian}
    \hat{H}=J\sum_{\langle ij\rangle}\hat{\boldsymbol{\sigma}}_i \cdot \hat{\boldsymbol{\sigma}}_j,
\end{equation}
where $(\hat{\sigma}_i^x, \hat{\sigma}_i^y, \hat{\sigma}_i^z)$ is the vector of the Pauli matrices describing spin at the site $i$ of the honeycomb lattice; $J=1$ is Heisenberg exchange interaction; sites $i$ and $j$ belong to the honeycomb lattice of $N=14$ sites depicted in the inset of Fig.~\ref{fig:figS1}(a); and $\langle ij\rangle$ signifies that the sum is over the nearest-neighbor sites. Unlike in the case of Kitaev QSL [Fig.~\ref{fig:Lindblad_results}], GMN remains non-zero in both non-loopy and loopy subregions at all times [Fig.~\ref{fig:figS1}(a)].

\begin{figure}[t!]
    \centering
    \includegraphics[width=0.9\linewidth]{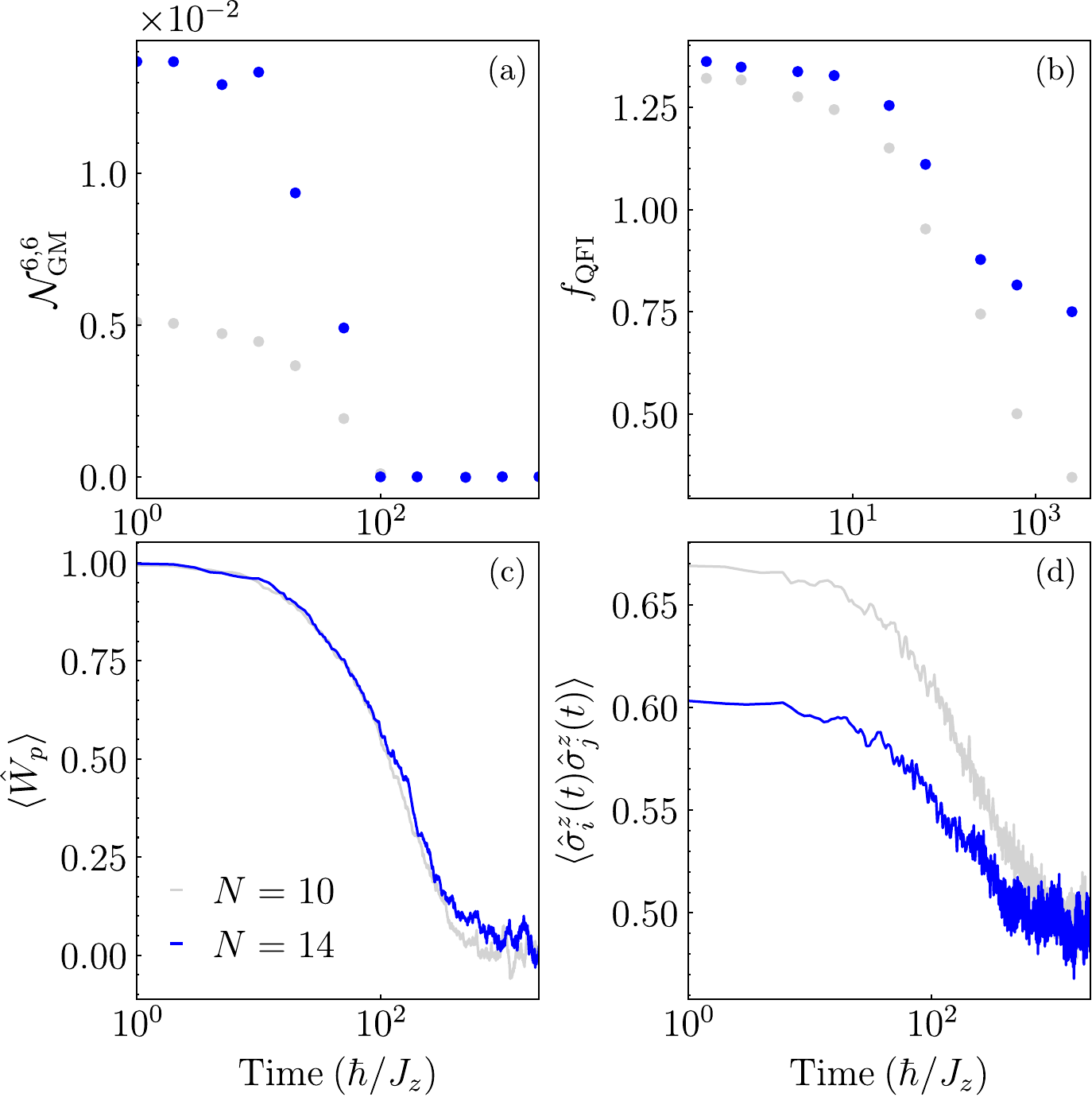}
    \caption{The same information as in Fig.~\ref{fig:Lindblad_results}   analyzing the Markovian regime, but for two different Kitaev QSL sizes. Note that blue curves here are identical to blue curves in Fig.~\ref{fig:Lindblad_results} for $N=14$ spins, while gray curves are for a smaller system of $N=10$ spins (comprising two hexagons). In both cases, the total system Kitaev QSL + bath is infinite with a continuous energy spectrum. The temperature of bosonic baths is $T=0.3 J_z$.}
    \label{fig:figS2}
\end{figure}


\section{Finite size effects} \label{sec:appendixb}

Figure~\ref{fig:figS2} gives the same information as Fig.~\ref{fig:Lindblad_results}, but using two different honeycomb lattice sizes, $N=10$ and $N=14$ (the latter size is used in Fig.~\ref{fig:Lindblad_results}), on which Kiteav QSL is defined. Finite size scaling is an important test~\cite{Laeuchli2019} to avoid artifacts in studies 
of closed quantum spin systems. For example, the limit of present classical digital computers, running the most sophisticated exact diagonalization algorithms, is $N=48$ spins $S=1/2$~\cite{Laeuchli2019}. Quantum computers could break this limit, but on present noisy intermediate-scale quantum hardware, a maximum of $N=12$ spins of the Kitaev QSL could be handled~\cite{Park2025}. Nevertheless, the very recent computation~\cite{Lyu2025} of GMN for closed QSLs has demonstrated that it becomes size-independent after $N=24$ spins are reached, which means  the thermodynamic limit is well-approximated by the largest system sizes (such as $N=36$) considered in that study. 

In the case of open quantum spin systems, Markovian dynamics via the {\tt QuTiP}~\cite{Johansson2013a, Lambert2024} package is limited to $N=14$ spins $S=1/2$, while non-Markovian dynamics via the \texttt{OQuPy} package~\cite{Fux2024} is limited to $N=21$ spins $S=1/2$. However, the whole system QSL + bath is very different from the closed QSL studied in Ref.~\cite{Lyu2025} as it is infinite with a continuous energy spectrum. Thus, the subsystem of quantum spins within the QSL + bath total system can often reach~\cite{Schuckert2018,ReyesOsorio2026} size-independent quantities at a rather small number of spins considered. In Fig.~\ref{fig:figS2} we demonstrate that changing the system size from $N=10$ to $N=14$ does not modify any  of our key findings regarding the time scale for the decay of four quantities computed.

\section{Effective Hamiltonian from reaction coordinate + polaron methodology}
\label{sec:appendixc}
The RC + polaron methodology could, in principle, be used~\cite{GarciaGaitan2024} to evolve many spins in the presence of a dissipative  non-Markovian environment while using the Lindblad QME. However, such evolution becomes computationally prohibitively expensive for $N=14$ spins we consider in Fig.~\ref{fig:TEMPO_results}. This is also the reason that in Fig.~\ref{fig:TEMPO_results} we time-evolve the open QSL via tensor network methodology~\cite{Fux2024}. Nevertheless, the RC + polaron methodology is useful for deriving an effective Hamiltonian in Eq.~\eqref{eq:effective_Hamiltonian} which describes the steady-state phase in the long-time limit of non-Markovian dynamics and assumes a low temperature of the bosonic bath.
The RC approach alone, initially formulated in Refs.~\cite{Anto-Sztrikacs2023, Min2024}, encounters obstacles when multiple baths are coupled to QSL via non-commuting operators. This is the case for both ``local coupling'' [Eq.~\eqref{eq:local}] and ``global coupling'' [Eq.~\eqref{eq:global}] cases. This problem can be solved~\cite{Garwola2024} by a non-factorizing polaron transformation, leading to an effective Hamiltonian 
\begin{equation}
    \label{eq:H_eff_detailed}
    \hat{H}_\mathrm{eff} = \hat{\Pi}_0 \hat{U}_P \hat{H}_S \hat{U}_P^\dagger \hat{\Pi}_0 -\sum_{i\mu} \frac{\lambda_i^2}{\Omega_i} (\hat{A}_{i\mu}^2)^\mathrm{eff},
\end{equation}
where $\hat{U}_P = \mathrm{exp} \left[ \sum_{i\mu} \lambda_i/\Omega_i (\hat{b}_i^\dagger -\hat{b}_i ) \hat{A}_{i\mu}\right]$ is the net polaron transformation. Here $\lambda_i$ is the $i$-th reaction coordinate coupling strength to the system;  $\Omega_i$ is its frequency; $\hat{A}_{i\mu}$ is the QSL operator coupled to the bath [Eqs.~\eqref{eq:local} and ~\eqref{eq:global}]; and effective operators $\hat{O}^\mathrm{eff}$ are defined as $\hat{O}^\mathrm{eff} = \hat{\Pi}_0 \hat{U_P} \hat{O}\hat{U}_P^\dagger \hat{\Pi}_0$. We computed  $\hat{O}^\mathrm{eff}$ by expressing operators of each reaction coordinate in the momentum representation~\cite{Garwola2024} as
\begin{equation}
    \label{eq:O_eff}
    \hat{O}^\mathrm{eff} = \int \hat{U}_P(\mathbf{p}) \hat{O} \hat{U}_P^\dagger(\mathbf{p}) \prod_n\frac{e^{-p_n^2}}{\sqrt{\pi}} dp_n,
\end{equation}
where $\mathbf{p}$ is a vector that gathers  all the momentum associated with different reaction coordinates, and $\hat{U}_P(\mathbf{p})=\exp \left( -i\sqrt{2} \sum_n p_n\lambda_n/\Omega_n \hat{A}_n\right)$. 

\subsubsection{Local coupling}

In the ``local coupling'' case [Eq.~\eqref{eq:local}], each spin is coupled to three identical baths via the $\hat{\sigma}_x$, $\hat{\sigma}_y$, $\hat{\sigma}_z$ operators, respectively. Note that the second term in Eq.~\eqref{eq:H_eff_detailed} will only contribute to a global shift in energy. Therefore, we focus on the first term by considering a generic Ising-like interaction in the Kitaev QSL Hamiltonian and its corresponding transformation. Since all RC degrees of freedom whose corresponding coupling to the system does commute with the term one is transforming vanish in Eq.~\eqref{eq:O_eff}, the transformed version of a generic Ising-like term will only involve six degrees of freedom (three per site) for the sites to which Ising-like interaction connects. The $\hat{U}_P (\mathbf{p})$ operators then become $4\times 4$ matrices that can be written as
\begin{equation}
    \hat{U}_P(\mathbf{p}) = f(\mathbf{p}) -ig(\mathbf{p}) \prod_{\mu,\nu \in\left\{ x,y,z\right\}} p_1^\mu p_2^\nu \hat{\sigma}_1^\mu \hat{\sigma}_2^\nu.
\end{equation}
Here $p_i = \sqrt{\sum_\mu (p_i^\mu)^2}$ is the norm; the subscript of the $p_1^i$ denotes the site to which RC is coupled, while its superscript denotes the spin operator to which it couples; $f(\mathbf{p}) = \cos (p_1 p_2 \lambda/\Omega)$; and $g(\mathbf{p}) = \sin(p_1 p_2 \lambda/\Omega)/(p_1p_2)$. Without the loss of generality, we consider the transformation of the term $\hat{\sigma}_1^z \hat{\sigma}_2^z$. To determine the integrand in Eq.~\eqref{eq:O_eff}, we use the following property of Pauli matrices, $\hat{\sigma}^\mu \hat{\sigma}^\nu = \delta_{\mu\nu} + i \epsilon_{\mu \nu \lambda}\hat{\sigma}^\lambda$, where $\delta_{\mu\nu}$ refers to the Kronecker delta, $\epsilon_{\mu\nu\lambda}$ is the Levi-Civita symbol, and the summation is assumed over repeated indices. After a lengthy but straightforward calculation, we obtained the following integral
\begin{widetext}
\begin{align}
    \label{eq:integrand}
    \hat{U}_P(\mathbf{p}) \hat{\sigma}_1^z \hat{\sigma}_2^z \hat{U}_P^\dagger(\mathbf{p})&=f^2(\mathbf{p}) \hat{\sigma}_1^z \hat{\sigma}_2^z+2f(\mathbf{p})g(\mathbf{p})\sum_{\mu\nu}\left[ p_1^z p_2^\mu \epsilon_{\mu z\nu} \hat{\sigma}_2^\nu + p_1^\mu p_2^z\epsilon_{\mu z\nu} \hat{\sigma}_1^\nu \right]\notag \\
    &+ g^2(\mathbf{p})\sum_{\mu\nu} \left[ \hat{\sigma}_1^\mu \hat{\sigma}_2^\nu (4p_1^\mu p_1^z p_2^\nu p_2^z) -2\hat{\sigma}_1^\mu \hat{\sigma}_2^z (p_1^\mu p_1^z (p_2^\nu)^2)-2\hat{\sigma}_1^z\hat{\sigma}_2^\nu((p_1^\mu)^2p_2^\nu p_2^z) + \hat{\sigma}_1^z \hat{\sigma}_2^z (p_1^\mu)^2 (p_2^\nu)^2\right].
\end{align}
\end{widetext}

The above expression can be further simplified by noting that both $f$ and $g$ functions are even. Since the integral domain is symmetric, the whole integral must be even to yield a non-zero value, so $\int \prod_n e^{-p_n^2}/\sqrt{\pi} g^2(\mathbf{p}) \sum_{\mu\nu} p_1^\mu p_1^z p_2^\nu p_2^z \propto \delta_{\mu z}\delta_{\nu z}$. Then, the integrand is reduced to
\begin{equation}
    \hat{U}_P(\mathbf{p}) \hat{\sigma}_1^z \hat{\sigma}_2^z \hat{U}_P^\dagger(\mathbf{p}) = \left[f^2(\mathbf{p})+g^2(\mathbf{p})(p_1^z)^2 (p_2^z)^2\right]\hat{\sigma}_1^z \hat{\sigma}_2^z,
\end{equation}
where the remaining six-dimensional integral is carried out by standard Monte Carlo integration and its dependence on the $\lambda/\Omega$ ratio is expressed in the scaling function $\kappa_{J_K}^\mathrm{loc}(\lambda/\Omega)$ in Eq.~\eqref{eq:Heff_local}.

\subsubsection{Global coupling}

In the ``global coupling'' case [Eq.~\eqref{eq:global}], three identical baths are coupled to each component of the total spin of the system, i.e., $\hat{A}_n = \sum_i \hat{\sigma}^n_i$. In this case, the second  term in Eq.~\eqref{eq:H_eff_detailed} becomes non-trivial because $\hat{A}_n^2 = NI +2\sum_{i<j}\hat{\sigma}_i^n \hat{\sigma}_j^n$. Just like in the ``local coupling,'' we focus on the transformation of $\hat{\sigma}_1^z \hat{\sigma}_2^z$, which can be simplified to the one involving $4\times 4$ matrices.  But in this case, the integral is three-dimensional. Then Eq.~\eqref{eq:integrand} is replaced by 
\begin{widetext}
\begin{equation}
    \hat{U}_P(\mathbf{p}) \hat{\sigma}_1^z \hat{\sigma}_2^z \hat{U}_P^\dagger(\mathbf{p}) = [f^2(\mathbf{p})+(-2(p_z)^2(p_x)^2 + 3(p_z)^4)g^2(\mathbf{p})] \hat{\sigma}_1^z \hat{\sigma}_2^z + g^2(\mathbf{p})4(p_z)^2(p_x)^2 \hat{\boldsymbol{\sigma}}_1 \cdot \hat{\boldsymbol{\sigma}}_2,
\end{equation}
\end{widetext}
where an additional Heisenberg interaction (last term on the right-hand side) is induced by the global nature of the bath. Here, we can simplify the momentum labels since no distinction of different sites is needed, so that the subscript of $p_\mu$ indicates the spin operator to which the corresponding RC is coupled.

\bibliography{biblio}


\end{document}